\documentclass[prd,twocolumn,showpacs,superscriptaddress]{revtex4-1}
\usepackage{lipsum}
\usepackage{hyperref}
\usepackage{graphicx}
\usepackage{graphics}
\usepackage{amssymb}
\usepackage{amsthm}
\usepackage{mathbbol}
\usepackage{color}
\usepackage{amsfonts}
\usepackage{booktabs}
\usepackage{mathtools}
\usepackage{marginnote}
\usepackage[normalem]{ulem}

\newcommand{\be}{\begin{equation}}
\newcommand{\ee}{\end{equation}}
\newcommand{\bea}{\begin{eqnarray}}
\newcommand{\eea}{\end{eqnarray}}
 
\newcommand{\gtapprox}{\raisebox{-0.5ex}{$\,\stackrel{>}{\scriptstyle\sim}\,$}}
\newcommand{\ltapprox}{\raisebox{-0.5ex}{$\,\stackrel{<}{\scriptstyle\sim}\,$}}
\begin{document}

\title{Lattice QCD investigation of the structure of the $a_0(980)$ meson}

\author{Constantia Alexandrou}
\affiliation{Department of Physics, University of Cyprus, P.O. Box 20537, 1678 Nicosia, Cyprus}
\affiliation{Computation-based Science and Technology Research Center, \\ The Cyprus Institute, 20 Kavafi Street, 2121 Nicosia, Cyprus}

\author{Joshua Berlin}
\affiliation{Goethe-Universit\"at Frankfurt am Main, Institut f\"ur Theoretische Physik, Max-von-Laue-Stra{\ss}e 1, D-60438 Frankfurt am Main, Germany}

\author{Mattia Dalla Brida}
\affiliation{Dipartimento di Fisica, Universit\`a di Milano-Bicocca and \\ INFN, Sezione di Milano-Bicocca, Piazza della Scienza 3, I-20126 Milano, Italy}

\author{Jacob Finkenrath}
\affiliation{Computation-based Science and Technology Research Center, The Cyprus Institute, 20 Kavafi Street, 2121 Nicosia, Cyprus}

\author{Theodoros Leontiou}
\affiliation{Department of Mechanical Engineering, Frederick University, 1036 Nicosia, Cyprus}

\author{Marc Wagner}
\affiliation{Goethe-Universit\"at Frankfurt am Main, Institut f\"ur Theoretische Physik, Max-von-Laue-Stra{\ss}e 1, D-60438 Frankfurt am Main, Germany}

\begin{abstract}
We investigate the quark content of the scalar meson $a_0(980)$ using lattice QCD. To this end we consider correlation functions of six different two- and four-quark interpolating fields. We evaluate all diagrams, including diagrams, where quarks propagate within a timeslice, e.g.\ with closed quark loops. We demonstrate that diagrams containing such closed quark loops have a drastic effect on the final results and, thus, may not be neglected. Our analysis shows that in addition to the expected spectrum of two-meson scattering states there is an additional energy level around the two-particle thresholds of $K + \bar{K}$ and $\eta + \pi$. This additional state, which is a candidate for the $a_0(980)$ meson, couples to a quark-antiquark as well as to a diquark-antidiquark interpolating field, indicating that it is a superposition of an ordinary $\bar{q} q$ and a tetraquark structure. The analysis is performed using AMIAS, a novel statistical method based on the sampling of all possible spectral decompositions of the considered correlation functions, as well as solving standard generalized eigenvalue problems.
\end{abstract}

\maketitle



\section{Introduction}

The mass ordering of the light scalar mesons $\sigma$, $\kappa$, $f_0(980)$ and $a_0(980)$, as observed in experiments, is inverted compared to what is expected based on the conventional quark-antiquark structure. A possible explanation of this mass ordering is to interpret these states as tetraquarks. Assuming such a four-quark structure the expected mass ordering is consistent with experimental results and the degeneracy of the $f_0(980)$ meson and the $a_0(980)$ meson is straightforward to understand \cite{Jaffe:2004ph}.

Several lattice QCD studies of light scalar mesons have been published in the last couple of years (cf.\ e.g.\ \cite{Bernard:2007qf,Gattringer:2008be,Prelovsek:2008qu,
Liu:2008ee,Prelovsek:2010kg,Wakayama:2012ne,
Prelovsek:2013ela,Wakayama:2014gpa,Dudek:2016cru}). Regarding the $a_0(980)$ meson the elaborate study presented in Ref.\ \cite{Dudek:2016cru} using L\"uscher's finite volume approach is of particular interest. The authors find a resonance close to the $K + \bar{K}$ threshold, which they interpret as the $a_0(980)$ meson. Moreover, cf.\ Ref.\ \cite{Guo:2016zep}, where the same lattice data has been analyzed using chiral effective field theory.

In this work we continue our lattice QCD investigation of the $a_0(980)$ meson \cite{Daldrop:2012sr,Alexandrou:2012rm,Wagner:2012ay,
Wagner:2013nta,Wagner:2013jda,Wagner:2013vaa,
Abdel-Rehim:2014zwa,Berlin:2015faa,Berlin:2016zci} with particular focus on clarifying its quark structure, i.e.\ whether it is of quark-antiquark type or rather a four-quark state. We include the computation of all diagrams, where quarks propagate within a timeslice, e.g.\ diagrams with closed quark loops, which were neglected in many previous investigations of scalar mesons. We show that these contributions are important and lead to the appearance of an additional $q\bar q$-like state around the two-particle thresholds of $\eta_s + \pi$ and $K + \bar{K}$, which could correspond to the $a_0(980)$ meson. On a technical level we apply the Athens Model Independent Analysis Scheme (AMIAS), an analysis method based on statistical concepts for extracting excited states from correlation functions. AMIAS is a novel analysis method, which has recently been used in a study of the nucleon
spectrum \cite{Alexandrou:2014mka}. AMIAS utilizes all the information encoded in the correlation function with the particular advantage of exploiting also data at small temporal separations, where statistical errors are typically small. In addition to AMIAS, we also use the standard generalized eigenvalue problem (GEVP) method, i.e.\ we solve a generalized eigenvalue problem and extract the spectrum from effective mass plateaus. Both GEVP and AMIAS provide information on the relative importance of the considered interpolating fields, which include quark-antiquark, four-quark and two-meson structures. Combining both methods allows to check the robustness of our results. 

This paper is organized as follows: In section~\ref{sec:simu} we describe the lattice techniques used and briefly discuss the inclusion and importance of diagrams, where quarks propagate within a timeslice. We demonstrate that these diagrams are essential both on a qualitative and quantitative level and, hence, may not be neglected. In section~\ref{sec:euccor} we present the spectral decomposition of two-point correlation functions and explain possible complications arising due to the presence of multi-particle states. A short description of AMIAS is provided in section~\ref{sec:amias}. In section~\ref{sec:cormat} we present the analysis of a $6\times 6$ correlation matrix containing various interpolating fileds using both GEVP and AMIAS. In section~\ref{sec:conclusions} we summarize our findings and give our conclusions.


\section{\label{sec:simu}Lattice setup and techniques}

To investigate the $a_0(980)$ meson, which has quantum numbers $I(J^P) = 1(0^+)$ and a mass around  $980 \, \textrm{MeV}$ \cite{Agashe:2014kda}, we consider a $6 \times 6$ correlation matrix
\begin{eqnarray}
C_{j k}(t) = \Big\langle \mathcal{O}^j(t_2) \mathcal{O}^{k \dagger}(t_1) \Big\rangle \quad , \quad t = t_2-t_1 .
\end{eqnarray}
The interpolating fields $\mathcal{O}^j$, $j=1,\ldots,6$ have either a two-quark $\bar{d} u$ or a four-quark $\bar{d} u \bar{s} s$ structure. The four-quarks can be  arranged as a meson-meson interpolating field (either a bound pair of mesons or two separated mesons) or in a diquark-antidiquark combination. We consider the interpolating fields
\begin{eqnarray}
\label{EQN002} & & \hspace{-0.7cm} \mathcal{O}^1 = \mathcal{O}^{q \bar{q}} = N_1 \sum_{\bf{x}} \Big({\bar d}({\bf x}) u({\bf x})\Big) \\
\nonumber & & \hspace{-0.7cm} \mathcal{O}^2 = \mathcal{O}^{K \bar{K}, \ \textrm{point}} \\
\label{EQN007} & & = N_2 \sum_{\bf{x}} \Big({\bar s}({\bf x}) \gamma_5 u({\bf x})\Big) \Big({\bar d}({\bf x}) \gamma_5 s({\bf x})\Big) \\
\nonumber & & \hspace{-0.7cm} \mathcal{O}^3 = \mathcal{O}^{\eta_{s} \pi, \ \textrm{point}} \\
\label{EQN117} & & = N_3 \sum_{\bf{x}} \Big({\bar s}({\bf x}) \gamma_5 s({\bf x})\Big) \Big({\bar d}({\bf x}) \gamma_5 u({\bf x})\Big) \\
\nonumber & & \hspace{-0.7cm} \mathcal{O}^4 = \mathcal{O}^{Q \bar{Q}} \\
\nonumber & & = N_4 \sum_{\bf{x}} \epsilon_{a b c} \Big({\bar s}_b({\bf x}) (C \gamma_5) {\bar d}_c^T({\bf x})\Big) \\
 & & \hspace{0.675cm} \epsilon_{a d e} \Big(u_d^T({\bf x}) (C \gamma_5) s_e({\bf x})\Big) \\
\nonumber & & \hspace{-0.7cm} \mathcal{O}^5 = \mathcal{O}^{K \bar{K}, \ \textrm{2part}} \\
\label{EQN113} & & = N_5 \sum_{{\bf x},{\bf y}} \Big({\bar s}({\bf x}) \gamma_5 u({\bf x})\Big) \Big({\bar d}({\bf y}) \gamma_5 s({\bf y})\Big) \\
\nonumber & & \hspace{-0.7cm} \mathcal{O}^6 = \mathcal{O}^{\eta_{s} \pi, \ \textrm{2part}} \\
\label{EQN003} & & = N_6 \sum_{{\bf x},{\bf y}} \Big({\bar s}({\bf x}) \gamma_5 s({\bf x})\Big) \Big({\bar d}({\bf y}) \gamma_5 u({\bf y})\Big) ,
\end{eqnarray}
where $C$ is the charge conjugation matrix. The normalization factors $N_j$ are chosen such that $C_{j j}(t=a) = 1$ (no sum over $j$), i.e.\ in a way that the six interpolating fields generate trial states with similar norm. All six interpolating fields couple to the $a_0(980)$ and other states with the same quantum numbers. For example, the interpolating fields $\mathcal{O}^5$ and $\mathcal{O}^6$ mostly generate the two-meson states $K + \bar{K}$ and $\eta + \pi$, respectively, which are expected to have masses close to the $a_0(980)$ state. In contrast to $\mathcal{O}^5$ and $\mathcal{O}^6$, where the two mesons both have zero momentum, the interpolating fields $\mathcal{O}^2$ and $\mathcal{O}^3$ represent two mesons at the same point in space, i.e.\ resemble mesonic molecules.

Previous results using these interpolating fields and Wilson clover fermions can be found in \cite{Abdel-Rehim:2014zwa,Berlin:2015faa,Berlin:2016zci}. In this work several significant improvements have been carried out:
\begin{itemize}
\item[(i)] we have improved the statistical accuracy of the correlation matrix $C_{j k}(t)$,

\item[(ii)] we include the propagation of strange quarks within a timeslice,

\item[(iii)] we consider all 36 elements of the correlation matrix,

\item[(iv)] we analyze the correlation matrix with both the standard GEVP and the AMIAS method.
\end{itemize}

We use an ensemble of around 500 gauge link configurations generated with $N_f=2+1$ dynamical Wilson clover quarks
and the Iwasaki gauge action produced by the PACS-CS Collaboration~\cite{Aoki:2008sm}. The lattice size is $64 \times 32^3$ with lattice spacing $a = 0.0907(14) \, \textrm{fm}$ and pion mass $m_\pi \approx 300 \, \textrm{MeV}$. To optimize the coupling of the interpolating fields to the low-lying energy eigenstates, quark fields are Gaussian smeared with APE smeared spatial gauge links (cf.\ e.g.\ \cite{Jansen:2008si} for detailed equations). 

\begin{figure}[htb]
\begin{center}
 \includegraphics[width=5.5cm]{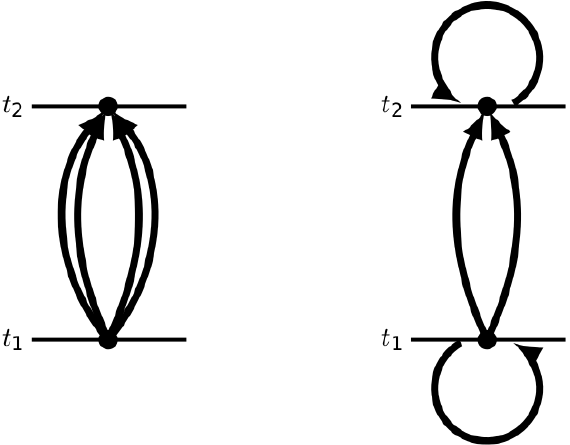}
\end{center}
\vskip-0.25cm
\caption{\label{fig:examplediagram}Diagrams forming the correlation function $C_{4 4}(t)$.}
\end{figure}

For each diagram of the correlation matrix $C_{jk}(t)$ we have implemented and compared various combinations of techniques including point-to-all propagators, stochastic timeslice-to-all propagators, the one-end trick and sequential propagators. Based on the results of these comparisons we have chosen for each diagram individually the most efficient combination of techniques for our computation of $C_{jk}(t)$ used in the physics analysis in section~\ref{sec:cormat}. Two example diagrams, which form the matrix element $C_{4 4}(t)$, are shown in Fig.~\ref{fig:examplediagram}. For the diagram on the left without closed quark loops we use four point-to-all propagators, i.e.\ its computation is technically simple and the statistical errors are quite small. Significantly more difficult is the diagram on the right with closed quark loops. In this case the use of four point-to-all propagators is not possible, because of the two closed quark loops. We found that using three point-to-all propagators and a stochastic timeslice-to-all propagator for one of the closed quark loops is the most efficient way of computing this particular diagram. For an extensive discussion we refer to \cite{Abdel-Rehim:2017dok}, where each diagram of the $6 \times 6$ correlation matrix is studied in detail. Finding efficient methods is particularly important for diagrams, where quarks propagate within a timeslice, e.g.\ diagrams containing closed quark loops. These diagrams are significantly more noisy than their counterparts, where quarks do not propagate within a timeslice: their noise-to-signal ratio grows exponentially with increasing temporal separation as discussed in \cite{Abdel-Rehim:2017dok}.

\begin{figure}[htb]
\begin{center}
\includegraphics[width=8.2cm]{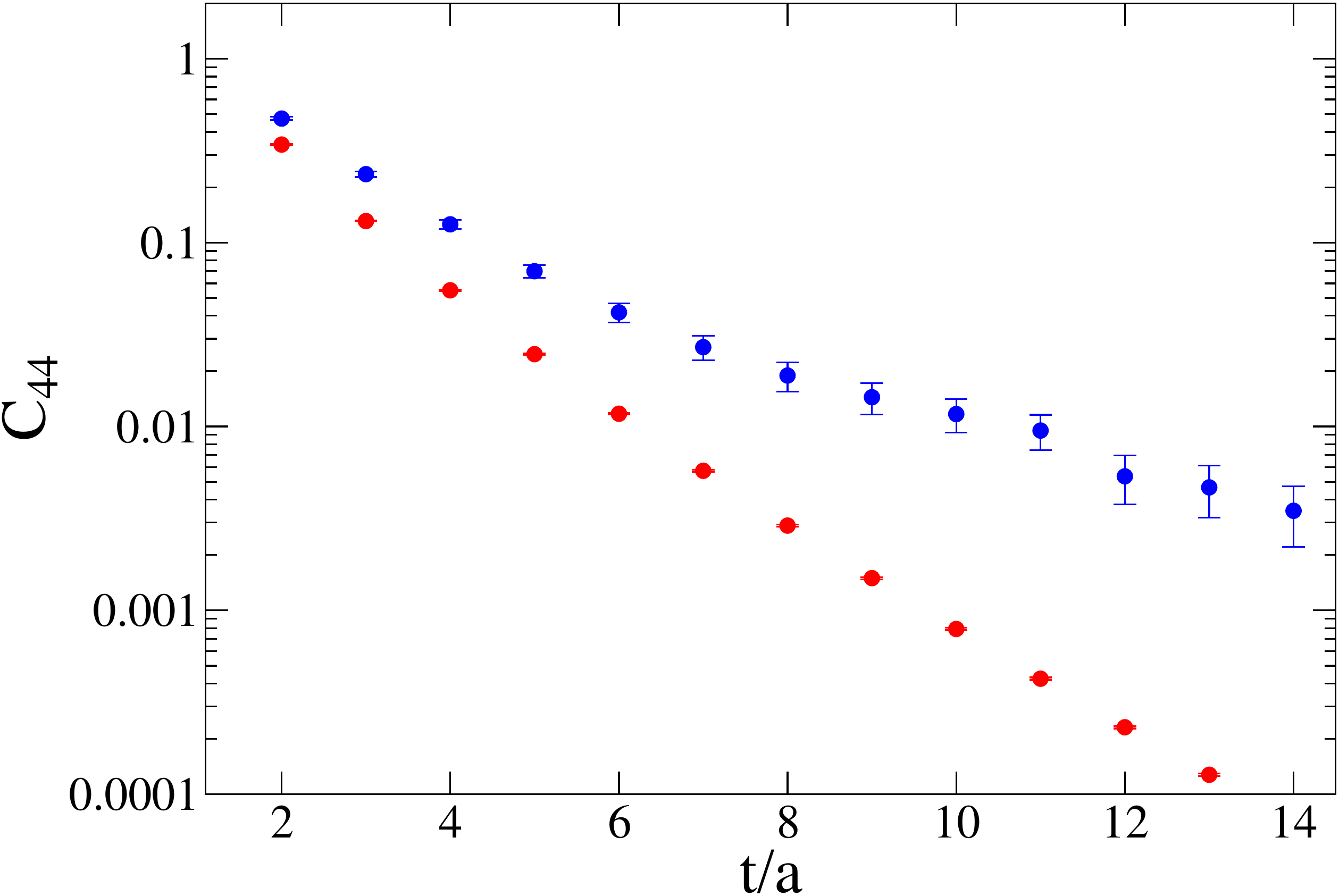}
\end{center}
\caption{\label{fig:effectofloops}$C_{4 4}(t)$ (blue points) compared to the diagram of $C_{4 4}(t)$, where quarks do not propagate within a timeslice (diagram on the left in Fig.\ \ref{fig:examplediagram}; red points).}
\end{figure}

Considering diagrams, where quarks propagte within a timeslice, is vital for any solid study of $a_0(980)$, because they lead to non-vanishing correlations between two-quark and four-quark interpolating fields, i.e.\ allow for $\bar{s} s$ creation and annihilation. Moreover, for correlation functions of two four-quark interpolating fields their contribution is sizable and should not be neglected as it has been done in the past, e.g.\ in \cite{Prelovsek:2010kg,Alexandrou:2012rm}. This is demonstrated by Fig.~\ref{fig:effectofloops}, where we show $C_{4 4}(t)$ both with (blue dots) and without (red dots) closed quark loops. Similar findings have been reported in Refs.\ \cite{Guo:2013nja,Wakayama:2014gpa}.

\section{\label{sec:euccor}Correlation functions on a periodic lattice}

A correlation function computed on a lattice with periodic temporal direction of extension $T$ can be expanded according to
\begin{eqnarray}
\nonumber & & \hspace{-0.7cm} C_{j k}(t) = \Big\langle \mathcal{O}^j(t) \mathcal{O}^{k \dagger}(0) \Big\rangle \\
\label{EQN566} & & = \frac{1}{Z} \sum_{m,n} e^{-E_m (T-t)} c_{m,n}^j e^{-E_n t} (c_{m,n}^k)^\ast
\end{eqnarray}
with energy eigenstates $| m \rangle$, corresponding energy eigenvalues $E_m$, $c_{m,n}^j = \langle m | \mathcal{O}^j | n \rangle$ and $Z = \sum_m e^{-E_m T}$. Using $u \leftrightarrow d$ flavor symmetry, i.e.\ $E_m = E_{m(u \leftrightarrow d)}$, where $m(u \leftrightarrow d)$ denotes the state $| m \rangle$ with $u$ flavor and $d$ flavor exchanged, one can rewrite Eq.\ (\ref{EQN566}) according to
\begin{eqnarray}
\nonumber & & \hspace{-0.7cm} C_{j k}(t) \\
\nonumber & & = \frac{1}{Z} \sum_{m,n} e^{-(E_m + E_n) T/2} \\
\nonumber & & \hspace{0.675cm} c^j_{m,n} (c^k_{m,n})^\ast e^{+E_{m,n} (t - T/2)} \\
\nonumber & & = \frac{1}{Z} \sum_{m,n} e^{-(E_m + E_n) T/2} \\
\nonumber & & \hspace{0.675cm} \frac{1}{2} \Big(c^j_{m,n} (c^k_{m,n})^\ast e^{+E_{m,n} (t - T/2)} \\
\nonumber & & \hspace{0.675cm} + c^j_{n(u \leftrightarrow d),m(u \leftrightarrow d)} (c^k_{n(u \leftrightarrow d),m(u \leftrightarrow d)})^\ast \\
\label{EQN563} & & \hspace{0.675cm}e^{-E_{m,n} (t - T/2)}\Big)
\end{eqnarray}
with $E_{m,n} = E_m - E_n$. For the interpolating fields defined by Eqs.\ (\ref{EQN002}) to (\ref{EQN003}) one can show
\begin{eqnarray}
c^j_{n(u \leftrightarrow d),m(u \leftrightarrow d)} (c^k_{n(u \leftrightarrow d),m(u \leftrightarrow d)})^\ast = c^j_{m,n} (c^k_{m,n})^\ast ,
\end{eqnarray}
which follows from
\begin{itemize}
\item $\mathcal{O}^j = \mathcal{O}_{u \leftrightarrow d}^{j \dagger}$ ($\mathcal{O}_{u \leftrightarrow d}^j$ denotes $\mathcal{O}^j$ with $u$ flavor and $d$ flavor exchanged),

\item $u \leftrightarrow d$ flavor symmetry,

\item $C_{j k}(t)$ is real (can be shown using standard symmetries of Wilson lattice QCD, e.g.\ $\gamma_5$ hermiticity).
\end{itemize}
Moreover, $c^j_{m,n} (c^k_{m,n})^\ast$ is real and without restriction one can choose real $c^j_{m,n}$, which we do in the following. Consequently, Eq.\ (\ref{EQN563}) can be simplified to
\begin{eqnarray}
\nonumber & & \hspace{-0.7cm} C_{j k}(t) = \frac{1}{Z} \sum_{m,n} e^{-(E_m + E_n) T/2} c^j_{m,n} c^k_{m,n} \\
\label{eq:cosh} & & \hspace{0.675cm} \textrm{cosh}\{E_{m,n} (t - T/2)\} .
\end{eqnarray}
Since the correlators are symmetric with respect to the reversal of time, $C_{j k}(t) = C_{j k}(T - t)$, it is sufficient to restrict the follwing discussion to temporal separations $0 \leq t \leq T/2$.

For $j = k$, for sufficiently large $T$, where $Z \approx e^{-E_\Omega T}$, and for sufficiently large $t$, Eq.\ (\ref{eq:cosh}) reduces to
\begin{eqnarray}
\nonumber & & \hspace{-0.7cm} C_{j j}(t) \\
\label{EQN001} & & \approx 2 (c^j_{\tilde{0},\Omega})^2 e^{-E_{\tilde{0},\Omega} T/2} \textrm{cosh}\{E_{\tilde{0},\Omega} (t - T/2)\} ,
\end{eqnarray}
if the correlation function is not contaminated by effects related to multi-particle states as discussed below ($| \Omega \rangle$ and $| \tilde{0} \rangle$ denote the vacuum and the lowest state in the $I(J^P) = 1(0^+)$ sector probed by the interpolating fields defined by Eqs.\ (\ref{EQN002}) to (\ref{EQN003}), respectively). Consequently, the energy difference $E_{\tilde{0},\Omega}$ can be extracted by fitting
\begin{eqnarray}
C_{j j}(t) = A \textrm{cosh}\{E_{\tilde{0},\Omega} (t - T/2)\}
\end{eqnarray}
to the lattice QCD results for the correlation function $C_{j j}(t)$ at sufficiently large $t$ with fitting parameters $E_{\tilde{0},\Omega}$ and $A$. Alternatively, one can solve the equation
\begin{eqnarray}
\label{eq:effEn} \frac{C_{j j}(t)}{C_{j j}(t-a)} = \frac{\textrm{cosh}\{E_\textrm{eff}(t) (t - T/2)\}}{\textrm{cosh}\{E_\textrm{eff}(t) (t-a - T/2)\}}
\end{eqnarray}
with respect to $E_\textrm{eff}(t)$, where $E_\textrm{eff}(t) \approx E_{\tilde{0},\Omega} = \textrm{const}$ for large $t$. In other words, a plateau-like behavior of $E_\textrm{eff}(t)$ indicates the mass $E_{\tilde{0},\Omega}$. 

A common method to extract several energy levels from an $N \times N$ correlation matrix is to solve the GEVP (cf.\ e.g.\ \cite{Blossier:2009kd} and references therein)
\begin{eqnarray}
\label{gevp2} C(t) \mathbf{v}_m(t,t_0) = \lambda_m(t,t_0) C(t_0) \mathbf{v}_m(t,t_0) ,
\end{eqnarray}
where $C(t)$ is the correlation matrix with entries $C_{j k}(t)$, $j,k=1,\ldots,N$, $\mathbf{v}_m(t,t_0)$ the eigenvector corresponding to the eigenvalue $\lambda_m(t,t_0)$ and $t_0 \geq a$ a parameter, where a typical choice is $t_0 = a$. $N$ effective energies $E_{\textrm{eff},m}(t)$ can be obtained by solving
\begin{eqnarray}
\label{eq:lambdaeffEn} \frac{\lambda_m(t,t_0)}{\lambda_m(t-a,t_0)} = \frac{\textrm{cosh}\{E_{\textrm{eff},m}(t) (t - T/2)\}}{\textrm{cosh}\{E_{\textrm{eff},m}(t) (t-a - T/2)\}}
\end{eqnarray}
for each eigenvalue $\lambda_m(t,t_0)$, $m=0,\ldots,N-1$. In practice, however, effective energies $E_{\textrm{eff},m}(t)$ often exhibit strong statistical fluctuations, in particular for large $t$ and $m > 0$, rendering a
reliable identification of plateaus and extraction of masses $E_{\tilde{m},\Omega}$ difficult ($| \tilde{0} \rangle, | \tilde{1} \rangle, \ldots$ denote the lowest states in the $I(J^P) = 1(0^+)$ sector probed by the interpolating fields defined by Eqs.\ (\ref{EQN002}) to (\ref{EQN003})). Therefore, besides using the GEVP, we also employ an alternative analysis approach, AMIAS, which is discussed in detail in section~\ref{sec:amias}.

\begin{figure}[htb]
\begin{center}
\includegraphics[width=8.2cm]{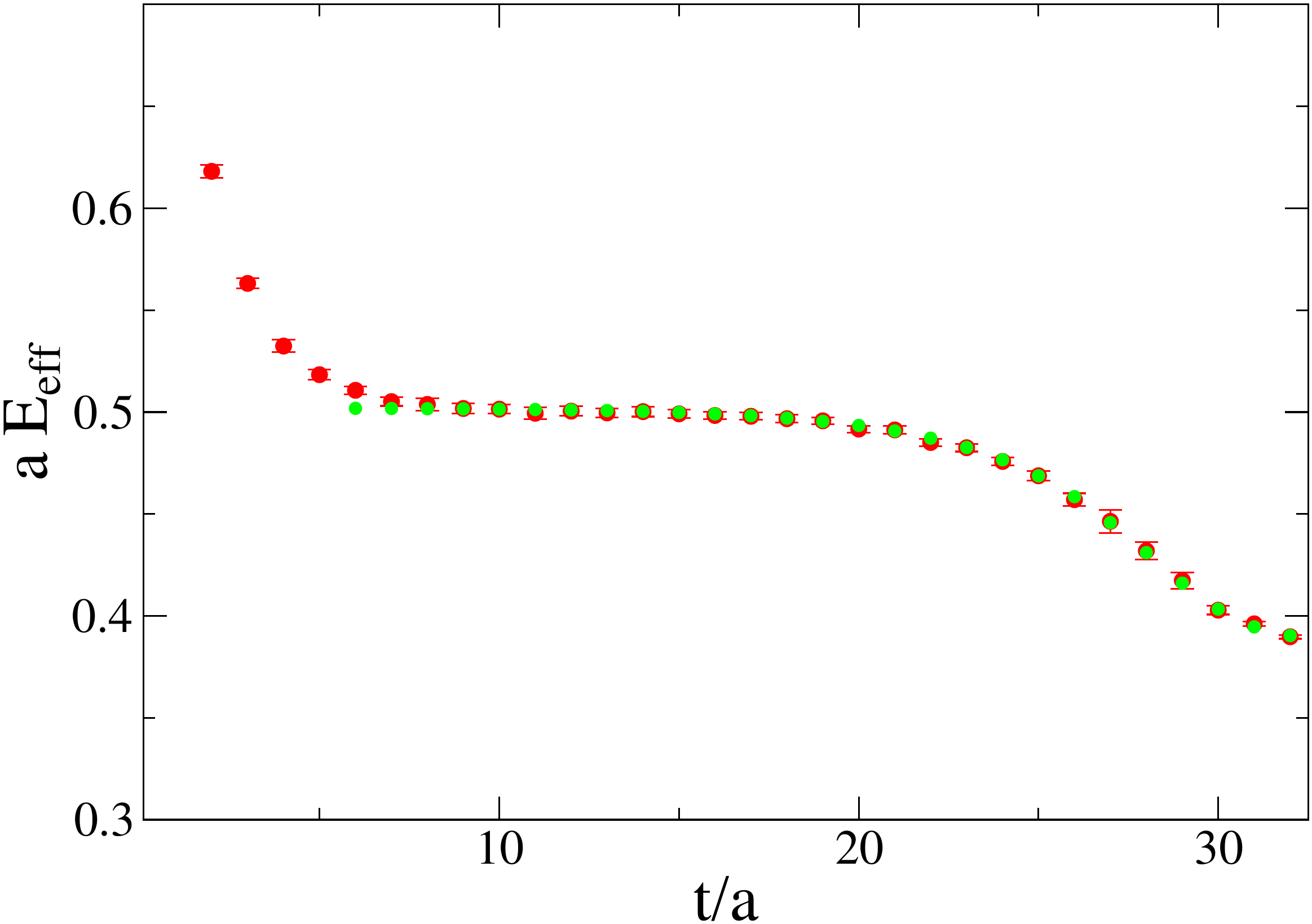}
\end{center}
\caption{\label{FIG001}$E_\textrm{eff}(t)$ according to Eq.\ (\ref{eq:effEn}) for the lattice QCD results for the correlation function $C_{6 6}(t)$ (red points; quark propagation within a timeslice neglected) and the analytical expectations (green points; Eq.\ \ref{EQN100}).}
\end{figure}

When low-lying multi-particle states are present in the investigated sector, the determination of masses becomes even more difficult. For example in the $I(J^P) = 1(0^+)$ sector the lowest state is expected to be a two-meson state composed of an $\eta$ meson and a $\pi$ meson. Clearly, the interpolating fields defined by Eqs.\ (\ref{EQN002}) to (\ref{EQN003}) do not only generate an $\eta + \pi$ state, when applied to the vacuum $| \Omega \rangle$, but also yield non-vanishing matrix elements $\langle \pi | \mathcal{O}^j | \eta \rangle$ and $\langle \eta | \mathcal{O}^j | \pi(u \leftrightarrow d) \rangle$, i.e.\ anihilate an $\eta$ meson and generate a $\pi$ meson and vice versa. Consequently, a significant contribution to $C_{j k}(t)$ is
\begin{eqnarray}
\nonumber & & \hspace{-0.7cm} \frac{2}{Z} e^{-(E_\pi + E_\eta) T/2} c^j_{\pi,\eta} c^k_{\pi,\eta} \textrm{cosh}\{E_{\pi,\eta} (t - T/2)\} \\
\nonumber & & \approx 2 e^{-(m_\pi + m_\eta) T/2} c^j_{\pi,\eta} c^k_{\pi,\eta} \\
\label{EQN620} & & \hspace{0.675cm} \textrm{cosh}\{(m_\eta - m_\pi) (t - T/2)\}
\end{eqnarray}
(cf.\ Eq.\ (\ref{eq:cosh})). Similarly, a term providing information about the mass of the $\eta + \pi$ state is
\begin{eqnarray}
\nonumber & & \hspace{-0.7cm} \frac{2}{Z} e^{-(E_{\eta + \pi} + E_\Omega) T/2} c^j_{\eta + \pi,\Omega} c^k_{\eta + \pi,\Omega} \\
\nonumber & & \hspace{0.675cm} \textrm{cosh}\{E_{\eta +\pi,\Omega} (t - T/2)\} \\
\nonumber & & \approx 2 e^{-m_{\eta + \pi} T/2} c^j_{\eta + \pi,\Omega} c^k_{\eta + \pi,\Omega} \\
\label{EQN621} & & \hspace{0.675cm} \textrm{cosh}\{m_{\eta + \pi} (t - T/2)\} .
\end{eqnarray}
Assuming coefficients $|c^j_{\pi,\eta}| \approx |c^j_{\eta + \pi,\Omega}|$ it is easy to see that in the region of $t \approx T/2$ the two terms are comparable in magnitude. Therefore, Eq.\ (\ref{EQN620}) needs to be taken into account, when extracting masses from the correlation matrix at large temporal separations. Only for sufficiently small temporal separations $t$ the contribution from Eq.\ (\ref{EQN620}) may be neglected, since the ratio of Eq.\ (\ref{EQN621}) and Eq.\ (\ref{EQN620}) grows exponentially $\propto e^{-2 m_\pi (t - T/2)}$ with decreasing $t$. This is illustrated in Fig.\ \ref{FIG001}, where we show the effective mass $E_\textrm{eff}(t)$ defined in Eq.\ (\ref{eq:effEn}) for lattice QCD results for the correlation function $C_{6 6}(t)$ corresponding to the interpolating field $\mathcal{O}^{\eta_{s} \pi, \ \textrm{2part}}$. Note that we have neglected closed quark loops and quark propagation within a timeslice for this computation, to obtain sufficiently precise results at large temporal separations. We also compare with the effective mass (\ref{eq:effEn}), where $C_{j j}(t)$ is the sum of Eqs.\ (\ref{EQN620}) and (\ref{EQN621}), the analytically expected dominating terms at large temporal separations $t$, i.e.\
\begin{eqnarray}
\nonumber & & \hspace{-0.7cm} C_{j j}(t) \propto \textrm{cosh}\{(m_\eta - m_\pi) (t - T/2)\} \\
\label{EQN100} & & + \textrm{cosh}\{m_{\eta + \pi} (t - T/2)\}
\end{eqnarray}
with $m_\eta a = 0.364$, $m_\pi a = 0.138$ (the $\eta$ meson and $\pi$ meson masses in our lattice setup) and $m_{\eta + \pi} = m_\eta + m_\pi$, and find excellent agreement as can be seen in Fig.\ \ref{FIG001}.

\section{\label{sec:amias}Basics of the AMIAS method}

In this section we summarize the basics of the AMIAS method \cite{Alexandrou:2008bp,Papanicolas:2012sb,Alexandrou:2014mka}. A more detailed description and application to the excited nucleon spectrum can be found in Ref.\ \cite{Alexandrou:2014mka}.

Lattice QCD results for correlation functions $C_{j k}(t)$ with $\mathcal{O}^j$, $j = 1,\ldots,6$ defined in Eq.\ (\ref{EQN002}) to Eq.\ (\ref{EQN003}) can be parameterized according to Eq.\ (\ref{eq:cosh}). After approximating $Z \approx e^{-E_\Omega T}$ and truncating the sums $\sum_{m,n}$ to a limited number of important terms, typically terms corresponding to small energy differences $|E_{m,n}|$, one obtains fit functions of the form
\begin{eqnarray}
\nonumber & & \hspace{-0.7cm} C_{j k}^{\textrm{fit}}(t) = \sum_{m,n}^\textrm{truncated} 2 e^{-(E_{m,\Omega} + E_{n,\Omega}) T/2} c^j_{m,n} c^k_{m,n} \\
\label{eq:cosh_fit} & & \hspace{0.675cm} \textrm{cosh}\{E_{m,n} (t - T/2)\} ,
\end{eqnarray}
which are appropriate for sufficiently large $t$ and $T$ (the factor $2$ is included, because of $u \leftrightarrow d$ flavor symmetry; states $m,n$ and states $n(u \leftrightarrow d),m(u \leftrightarrow d)$ contribute with identical terms and, thus, should be combined in the fit functions). In practice it is convenient to use the equivalent fit functions
\begin{eqnarray}
\label{EQN006} C_{j k}^{\textrm{fit}}(t) = \sum_X^\textrm{truncated} a^j_X a^k_X \textrm{cosh}\{\mathcal{E}_X (t - T/2)\}
\end{eqnarray}
instead of Eq.\ (\ref{eq:cosh_fit}), where $X$ is a superindex replacing $\{ m,n \}$, $a_X \equiv \sqrt{2} e^{-(E_{m,\Omega} + E_{n,\Omega}) T/4} c^j_{m,n}$ and $\mathcal{E}_X \equiv E_{m,n}$. The fit parameters $E_X$ and $a^j_X$ are real. In the following these fit parameters are collectively denoted by $\mathcal{A}_r$.

AMIAS determines a probability distribution function (PDF) $\Pi(\mathcal{A}_r)$ for each fit parameter $\mathcal{A}_r$. The estimates for the values of the fit parameters and their uncertainties are the expectation values and the standard deviations of the corresponding PDFs,
\begin{eqnarray}
 & & \hspace{-0.7cm} \overline{\mathcal{A}}_r = \int d\mathcal{A}_r \, \mathcal{A}_r \Pi(\mathcal{A}_r) \\
 & & \hspace{-0.7cm} \sigma(\mathcal{A}_r) = \bigg(\int d\mathcal{A}_r \, (\mathcal{A}_r - \overline{\mathcal{A}}_r)^2 \Pi(\mathcal{A}_r)\bigg)^{1/2} .
\end{eqnarray}
AMIAS is able to handle a rather large number of parameters using Monte Carlo techniques, i.e.\ it is suited to study several energy eigenstates, if the lattice QCD results for correlation functions are sufficiently precise.

The PDF for the complete set of fit parameters is defined by
\begin{eqnarray}
\label{EQN004} P(\mathcal{A}_1,\mathcal{A}_2,\ldots) = \frac{1}{N} e^{-\chi^2 / 2}
\end{eqnarray}
with appropriate normalization $N$ and
\begin{eqnarray}
\label{eq:chi2} \chi^2 = \sum_{j,k} \sum_{t=t_{\textrm{min}}}^{t_{\textrm{max}}} \frac{(C_{j k}(t)- C_{j k}^{\textrm{fit}}(t))^2}{(\sigma_{j k}(t))^2} , 
\end{eqnarray}
which is the well-known $\chi^2$ used e.g.\ in $\chi^2$ minimizing fits. $C_{j k}(t)$ denotes 
the correlation functions computed using lattice QCD with corresponding statistical errors $\sigma_{j k}(t)$, while $C_{j,k}^{\textrm{fit}}(t)$ is given by Eq.\ (\ref{eq:cosh_fit}). To obtain the PDF $\Pi(\mathcal{A}_r)$ for a specific fit parameter $\mathcal{A}_r$, one has to integrate Eq.\ (\ref{EQN004}) over all other parameters. In particular, the probability for the parameter $\mathcal{A}_r$ to be inside the interval $[a,b]$ is
\begin{eqnarray}
\nonumber & & \hspace{-0.7cm} \int_a^b d\mathcal{A}_r \, \Pi(\mathcal{A}_r) = \frac{
          \int_a^b d\mathcal{A}_r \,
          \int_{-\infty}^{+\infty} \prod_{s \ne r} d\mathcal{A}_s \,
          e^{-\chi^2/2}
          }
          {
          \int_{-\infty}^{+\infty} \prod_s d\mathcal{A}_s \,
          e^{-\chi^2/2}
          } . \\
\label{eq:montecarlo} & &
\end{eqnarray}
This multi-dimensional integral can be computed with standard Monte Carlo methods. We implemented a par\-allel-tempering scheme combined with the Metropolis algorithm as described in detail in Ref.\ \cite{Alexandrou:2014mka}. The parallel-tempering scheme prevents the algorithm from getting stuck in a region around a local minimum of
$\chi^2$ and guarantees ergodicity of the algorithm.

A common choice for $t_\textrm{min}$ and $t_\textrm{max}$ in Eq.\ (\ref{eq:chi2}) is $t_\textrm{min} = a$ and $t_\textrm{max} = T/2$. With AMIAS one can determine the maximum number of parameters, to which the lattice QCD results for the correlation functions are sensitive on, i.e.\ the number of terms considered in the truncated sum in Eq.\ (\ref{EQN006}). The strategy is to increase the number of fit parameters, until there is no observable change in the PDFs for the low-lying energy eigenstates of interest. For a detailed example cf.\ Ref.\ \cite{Alexandrou:2014mka}. For correlation functions with large statistical errors it might be helpful to also vary $t_\textrm{min}$ and $t_\textrm{max}$ and check the stability of the results, as e.g.\ done in section~\ref{EQN856}. 

\begin{figure}[htb]
\begin{center}
\includegraphics[width=8.2cm]{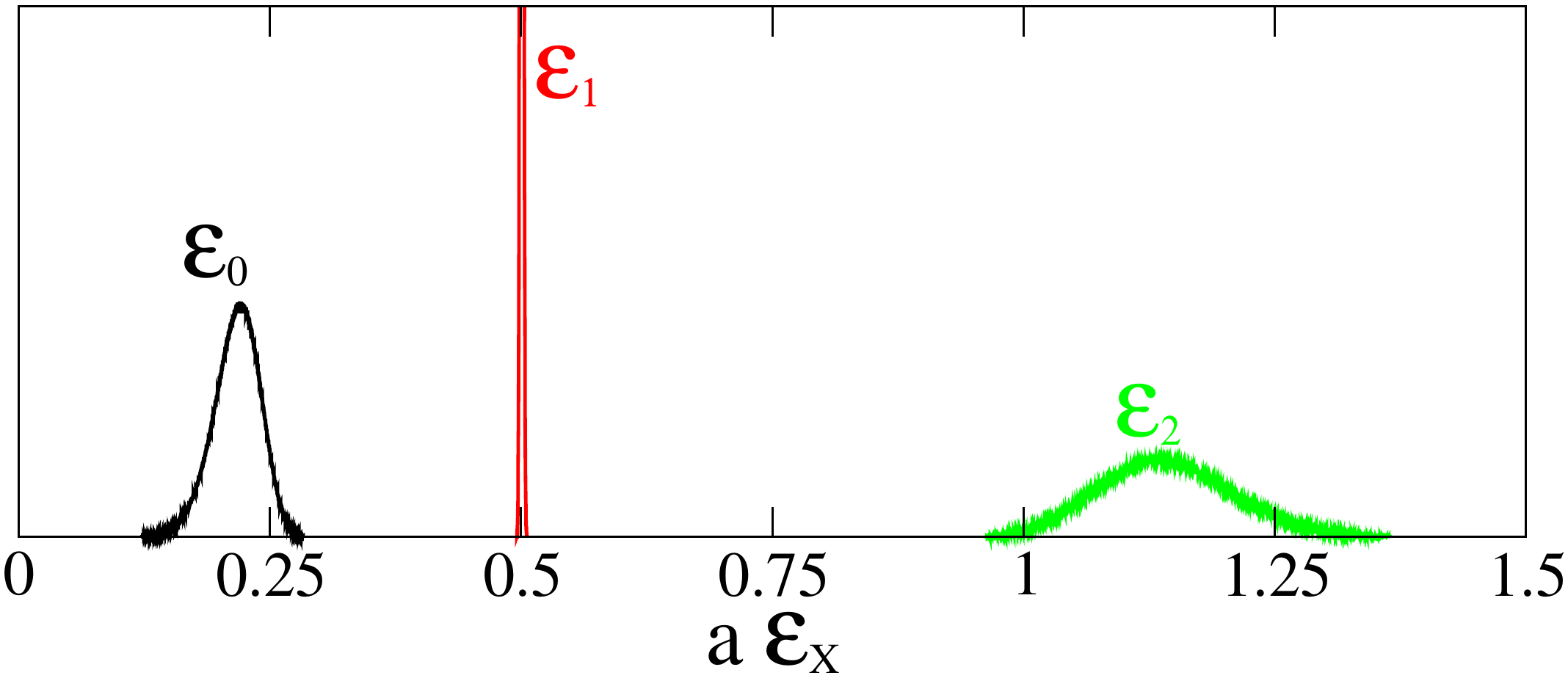}
\end{center}
\caption{\label{FIG002}AMIAS analysis of the correlation function $C_{6 6}(t)$, PDFs for the parameters $\mathcal{E}_0$, $\mathcal{E}_1$, $\mathcal{E}_2$ using three terms in the truncated sum of the fit function (\ref{eq:cosh_fit}) (quark propagation within a timeslice neglected).}
\end{figure}

As an example we show in Fig.\ \ref{FIG002} the AMIAS analysis of the correlation function $C_{6 6}(t)$ corresponding to the interpolating field $\mathcal{O}^{\eta_{s} \pi, \ \textrm{2part}}$, which we already discussed at the end of section~\ref{sec:euccor}. We have found that using $t_\textrm{min} = a$, $t_\textrm{max} = T / 2 = 32 a$ and three terms in the truncated sum of the fit function (Eq.\ (\ref{eq:cosh_fit})) allows to determine two energy differences rather precisely, $\mathcal{E}_0$ corresponding to $E_{\eta,\pi} a = (E_\eta - E_\pi) a \approx 0.23$ and $\mathcal{E}_1$ corresponding to $E_{\eta + \pi,\Omega} = (E_{\eta + \pi} - E_{\Omega}) a \approx m_\eta + m_\pi \approx 0.50$. $\mathcal{E}_2$ should not be interpreted as a specific energy difference, since it is unstable under a variation of the number of terms in the truncated sum.

\section{\label{sec:cormat}Analysis of the correlation matrix and numerical results for the $a_0(980)$ meson}


\subsection{Extraction of energy differences and amplitudes}

In this section we analyse the $6 \times 6$ correlation matrix discussed in section~\ref{sec:euccor} and various submatrices using both the GEVP method and the AMIAS method. The latter has proven to be particularly suited to study excited states \cite{Alexandrou:2014mka}. Both methods yield consistent results, which we consider to be an important cross-check, in particular due to the fact that the signal-to-noise ratios of the elements of the correlation matrix grow rapidly with increasing temporal separations.

To extract energy differences $E_{m,n}$ with the GEVP method, effective energies are computed as defined in Eq.\ (\ref{eq:lambdaeffEn}) (we always use $t_0 = a$ in Eq.\ (\ref{gevp2})). The plateau values at sufficiently large temporal separation correspond to $E_{m,n}$ as determined by fitting constants. The same energy differences $E_{m,n}$ are computed with the AMIAS method using the fit function (\ref{EQN006}) as explained in detail in section~\ref{sec:amias}.

The components of the eigenvectors $\mathbf{v}_m(t,t_0)$ obtained by solving a GEVP (\ref{gevp2}) provide information about the structure of the corresponding energy eigenstates:
\begin{equation}
\label{amiascoefs} | m \rangle \approx \sum_j v^j_m(t,t_0) \mathcal{O}^{j \dagger} | \Omega \rangle ,
\end{equation}
for sufficiently large $t$, where the approximate equality sign denotes the expansion of the energy eigenstate $| m \rangle$ within the subspace generated by the interpolating fields via $\mathcal{O}^{j \dagger} | \Omega \rangle$. We always normalize the eigenvectors according to $(\mathbf{v}_m(t,t_0))^2 = 1$, before plotting them.

One can easily convert the amplitudes $a_X^j$ extracted with AMIAS to $c^j_{\Omega,m}$ introduced in section~\ref{sec:euccor} via
\begin{eqnarray}
c^j_{m,n} = a_X / \sqrt{2} e^{-(E_{m,\Omega} + E_{n,\Omega}) T/4} .
\end{eqnarray}
Since $c^j_{\Omega,m} = \langle m | \mathcal{O}^{j \dagger} | \Omega \rangle$, these are the coefficients of the expansions of the trial states $\mathcal{O}^{j \dagger} | \Omega \rangle$ in terms of the energy eigenstates $| m \rangle$, i.e.\
\begin{eqnarray}
\label{EQN630} \mathcal{O}^{j \dagger} | \Omega \rangle \approx \sum_m^\textrm{truncated} c^j_{\Omega,m} | m \rangle .
\end{eqnarray}
More interesting, however, is invertig Eq.\ (\ref{EQN630}) and writing the extracted energy eigenstates in terms of the trial states,
\begin{eqnarray} \label{amiascoefs}
| m \rangle \approx \sum_j \tilde{v}^j_m \mathcal{O}^{j \dagger} | \Omega \rangle .
\end{eqnarray}
One can show that the matrix formed by the coefficients $\tilde{v}^j_m$ is the inverse of the matrix formed by the coefficients $c^j_{\Omega,m}$ up to exponentially small corrections, i.e.\
\begin{eqnarray}
\label{EQN245} \sum_j \tilde{v}^j_m c^j_{\Omega,n} \approx \delta_{m,n} .
\end{eqnarray}
Note that the coefficients $\tilde{v}^j_m$ are equivalent to the eigenvector components $v^j_m(t,t_0)$ obtained by solving a GEVP (\ref{gevp2}) and, thus, the resuls from the two methods can be compared in a meaningful way, after choosing the same normalization $(\tilde{\mathbf{v}}_m)^2 = 1$.


\subsection{Neglecting quark propagation within a timeslice}

At first we neglect diagrams, where quarks propagate within a timeslice. Thus, $\bar{s} s$ pair creation and annihilation is excluded. Consequently, the quark-antiquark interpolating field $\mathcal{O}^1$ probes a different sector than the four-quark interpolating fields $\mathcal{O}^2$ to $\mathcal{O}^6$. Thus, within this subsection we restrict the analysis to the $5 \times 5$ correlation matrix formed by the interpolating fields $\mathcal{O}^2$ to $\mathcal{O}^6$. Neglecting quark propagation within a timeslice leads to results with rather small statistical uncertainties and, thus, allows to cross-check our analysis methods and to compare with our previous study of the $a_0(980)$ meson \cite{Alexandrou:2012rm}, where we used a different lattice discretization and setup. Note, however, that contributions from diagrams, where quarks propagate within a timeslice, are sizeable (cf.\ e.g.\ Fig.\ \ref{fig:examplediagram} and \cite{Wakayama:2014gpa,Abdel-Rehim:2017dok}) and have to be taken into account to arrive at full QCD results, which can be compared to experimental data in a meaningful way.

\begin{figure}[htb]
\begin{center}
\includegraphics[width=8.2cm]{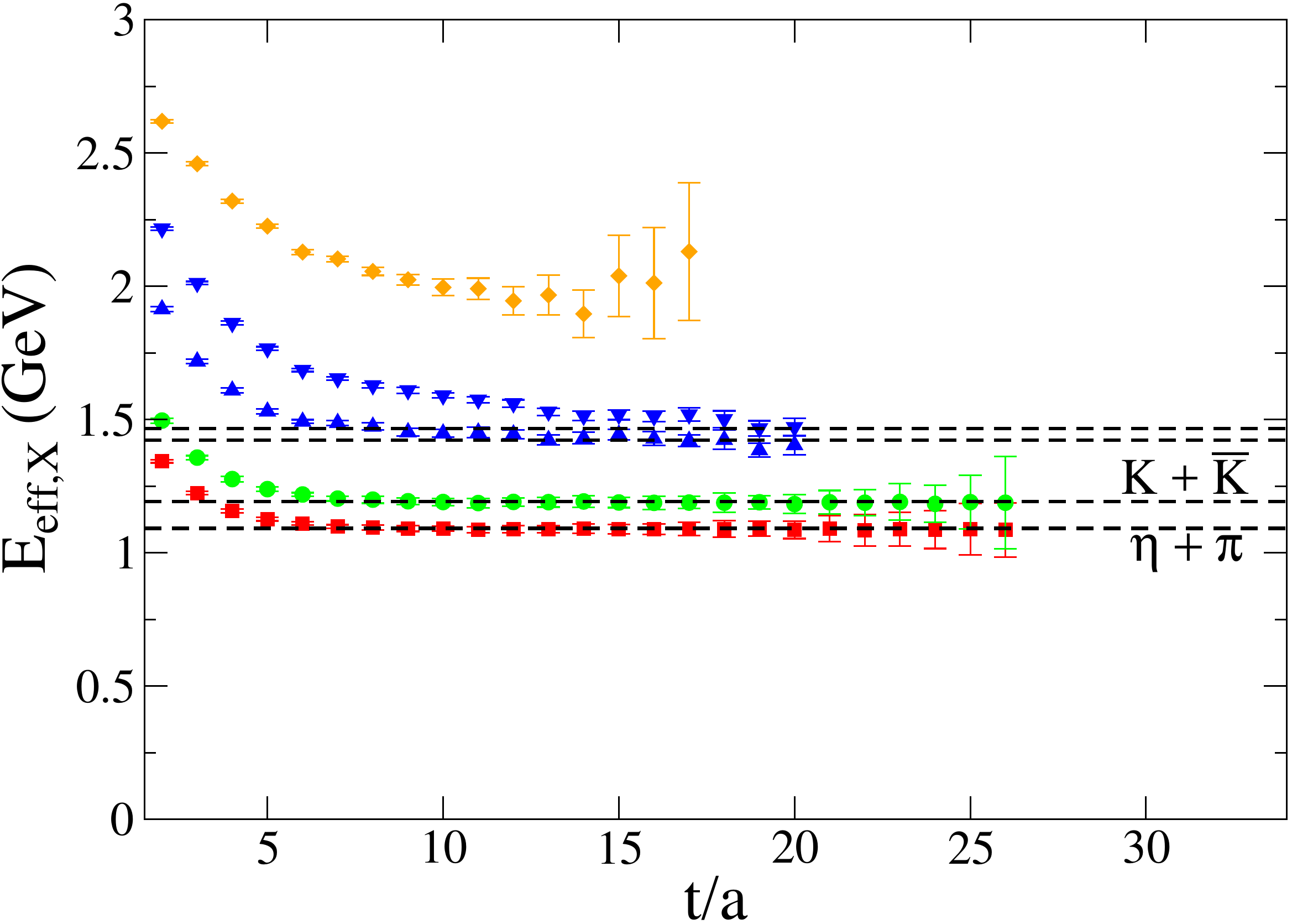}
\end{center}
\caption{\label{FIG005}Effective energies obtained with the GEVP method (quark propagation within a timeslice neglected; $5 \times 5$ correlation matrix with interpolating fields defined by Eqs.\ (\ref{EQN007}) to (\ref{EQN003})).}
\end{figure}

In Fig.\ \ref{FIG005} we show effective energies obtained by solving the GEVP for the $5 \times 5$ correlation matrix. In the absence of quark propagation within a timeslice all five effective energies exhibit convincing plateaus and the corresponding energy differences can be determined in a straightforward and precise way by fitting constants at large temporal separations, e.g.\ at $t \gtapprox 10 \, a \ldots 15 \, a$. The plateaus are consistent with the two-particle thresholds
\begin{itemize}
\item $m_\eta + m_\pi \approx 1092 \, \textrm{MeV}$,

\item $2 m_K \approx 1192 \, \textrm{MeV}$
\end{itemize}
as shown in Fig.\ \ref{FIG005} (note that the light quarks are unphysically heavy, corresponding to $m_\pi \approx 300 \, \textrm{MeV}$, $m_K \approx 596 \, \textrm{MeV}$ and $m_\eta \approx 792 \, \textrm{MeV}$). The lowest momentum excitations are given by
\begin{itemize}
\item $(m_\eta^2 + p_\textrm{min}^2)^{1/2} + (m_\pi^2 + p_\textrm{min}^2)^{1/2} \approx 1422 \, \textrm{MeV}$,

\item $2 (m_K^2 + p_\textrm{min}^2)^{1/2} \approx 1466 \, \textrm{MeV}$,
\end{itemize}
where $p_\textrm{min} = 2 \pi / L$ denotes one quantum of momentum and $L$ is the spatial lattice extent. We do not observe any sign of an additional energy level near the expected mass of the $a_0(980)$ meson, i.e.\ in the region around $1100 \, \textrm{MeV}$ to $1200 \, \textrm{MeV}$, which could be interpreted as the $a_0(980)$ meson \footnote{The mass of $a_0(980)$ is around two times the mass of $K$. Since in our lattice QCD setup $m_K \approx 596 \, \textrm{MeV}$ is unphysically heavy due to the unphysically heavy $u$ and $d$ quark mass, also the expected mass of $a_0(980)$ is larger than the corresponding experimental result $m_{a_0(980)} \approx 980 \, \textrm{MeV}$}.











\begin{figure}[htb]
\begin{center}
\includegraphics[width=8.2cm]{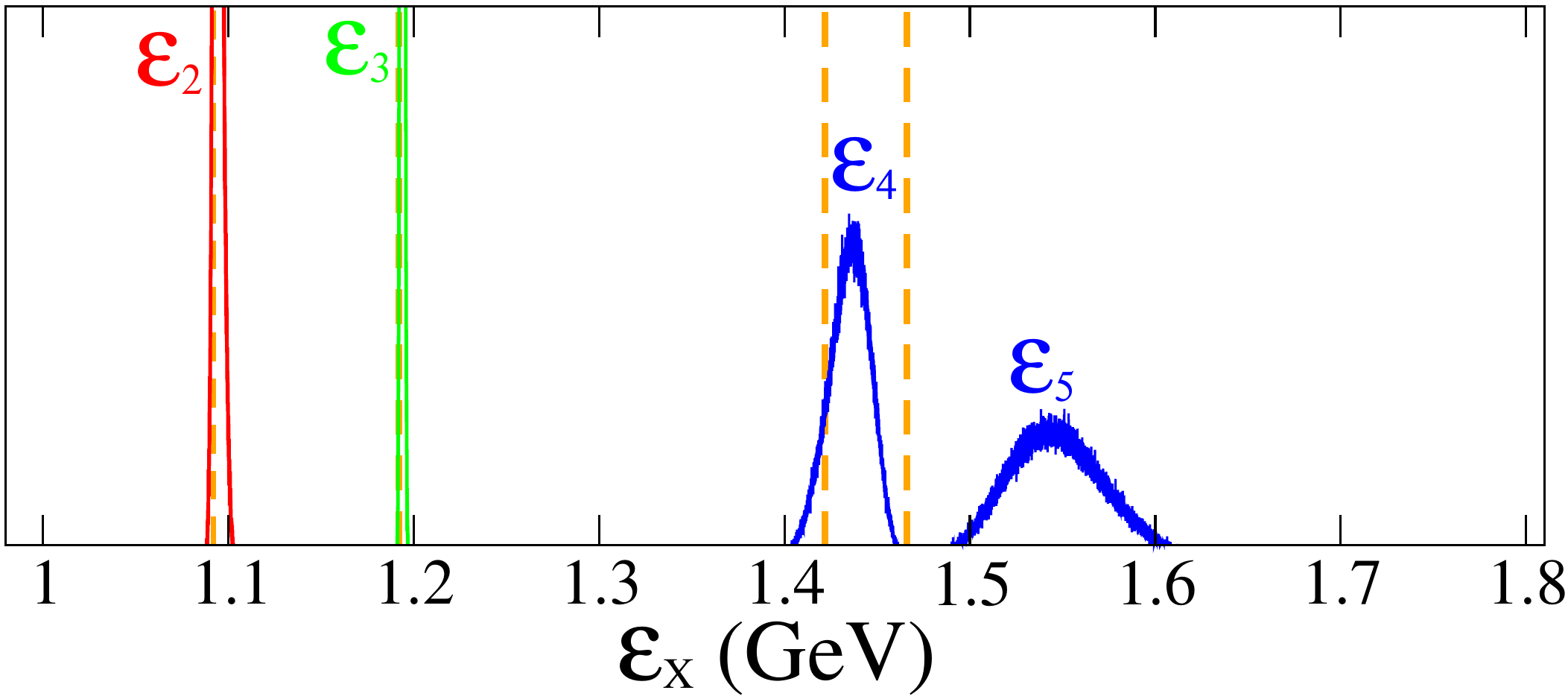} \\
\vspace{0.2cm}
\includegraphics[width=4.1cm,page=1]{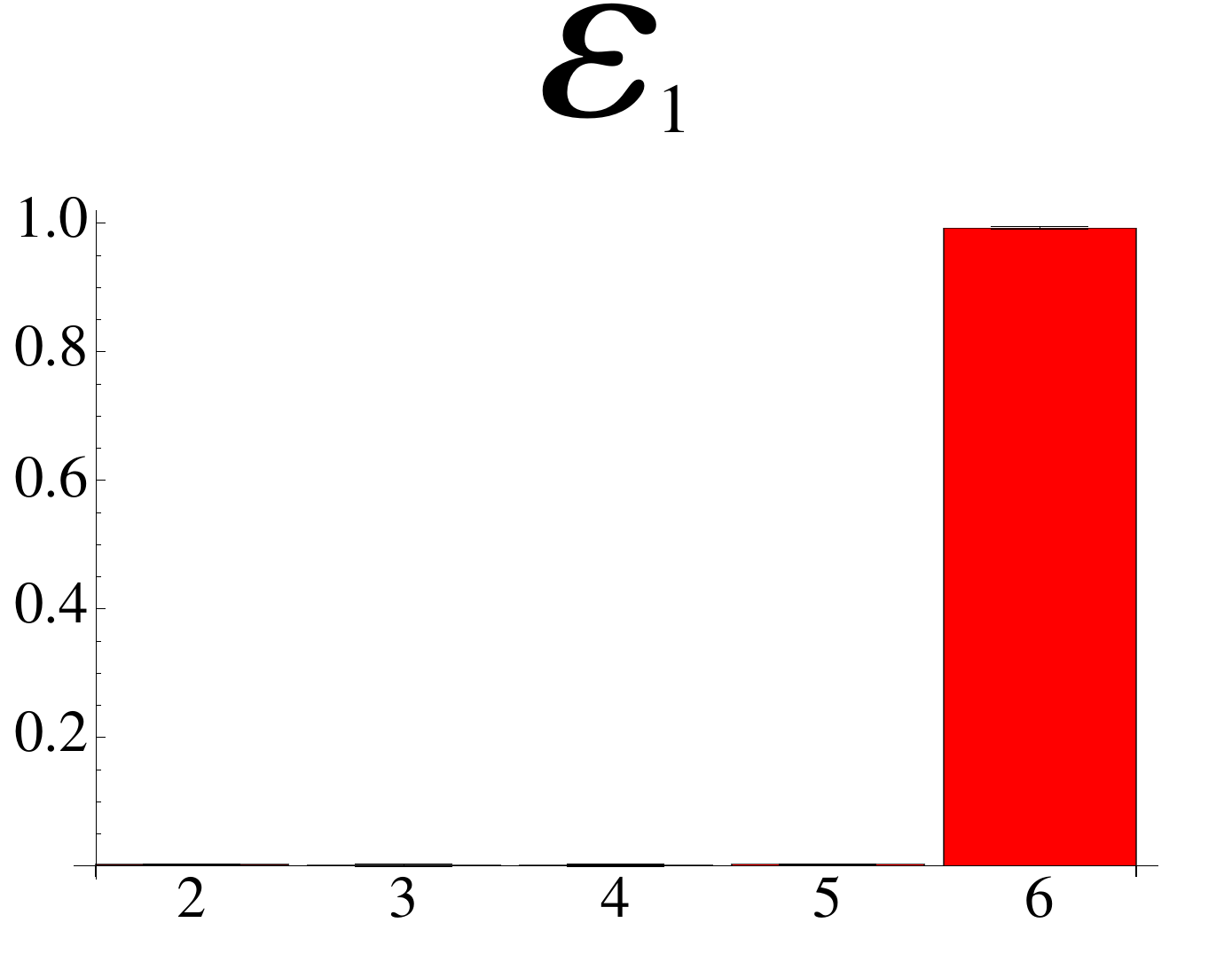}
\includegraphics[width=4.1cm,page=2]{figures/fig06b.pdf} \\
\includegraphics[width=4.1cm,page=3]{figures/fig06b.pdf}
\includegraphics[width=4.1cm,page=4]{figures/fig06b.pdf}
\end{center}
\caption{\label{fig:AMIAS}PDFs for energy differences and corresponding coefficients $\tilde{v}^j_m$ obtained with the AMIAS method (quark propagation within a timeslice neglected; $5 \times 5$ correlation matrix with interpolating fields defined by Eqs.\ (\ref{EQN007}) to (\ref{EQN003})).}
\end{figure}

The same energy differences are found, when using the AMIAS method. The corresponding PDFs, generated with the fit function given in Eq.\ (\ref{EQN006}) and eight terms in the truncated sum \footnote{Eight terms in the truncated sum lead to stable results for the energy differences $\mathcal{E}_2$ to $\mathcal{E}_5$, i.e.\ there is no significant change in the corresponding PDFs, when using nine or ten terms.}, is shown in Fig.\ \ref{fig:AMIAS}. To generate quantitative results including uncertainties we compute the mean values and widths of the PDFs. We find perfect agreement with the expected energies of states with two particles at rest,
\begin{itemize}
\item $\mathcal{E}_2 = 1095(2) \, \textrm{MeV}$ \\ (expectation: $m_\pi + m_\eta \approx 1092 \, \textrm{MeV}$),

\item $\mathcal{E}_3 = 1194(9) \, \textrm{MeV}$ \\ (expectation: $2 m_K \approx 1192 \, \textrm{MeV}$),
\end{itemize}
and fair agreement, when these particles have one quantum of relative momentum,
\begin{itemize}
\item $\mathcal{E}_4 = 1435(12) \, \textrm{MeV}$ \\ (expectation: \\ $(m_\eta^2 + p_\textrm{min}^2)^{1/2} + (m_\pi^2 + p_\textrm{min}^2)^{1/2} \approx 1422 \, \textrm{MeV}$),

\item $\mathcal{E}_5 = 1548(29) \, \textrm{MeV}$ \\ (expectation: $2 (m_K^2 + p_\textrm{min}^2)^{1/2} \approx 1466 \, \textrm{MeV}$).
\end{itemize}
When including correlation matrix data for temporal separations around $t = T/2$, which has small statistical uncertainties, when quark propagation within a timeslice is neglected, AMIAS also finds two significantly smaller energy differences $\mathcal{E}_0$ and $\mathcal{E}_1$ in the region of $m_K - m_{\bar{K}} = 0$ and $m_\eta - m_\pi$. This is expected and has been discussed in detail in section~\ref{sec:euccor}.

To support our interpretation of the states corresponding to $\mathcal{E}_m$, $m = 2,\ldots,5$, we show the corresponding coefficients $\tilde{v}^j_m$ in Fig.\ \ref{fig:AMIAS}. The states corresponding to $\mathcal{E}_2$ and $\mathcal{E}_3$ are clearly two-particle states $\eta + \pi $ and $K + \bar{K}$ with both mesons at rest, since the coefficients $\tilde{v}^j_m$ indicate a strong domination of interpolating fields $\mathcal{O}^5 = \mathcal{O}^{K \bar{K}, \ \textrm{2part}}$ and $\mathcal{O}^6 = \mathcal{O}^{\eta_{s} \pi, \ \textrm{2part}}$. The states corresponding to $\mathcal{E}_4$ and $\mathcal{E}_5$ exhibit significant contributions from interpolating fields $\mathcal{O}^2 = \mathcal{O}^{K \bar{K}, \ \textrm{point}}$ and $\mathcal{O}^3 = \mathcal{O}^{\eta_{s} \pi, \ \textrm{point}}$, which is consistent with their interpretation as two-particle states with one quantum of relative momentum. A similar GEVP analysis \cite{Berlin:2016zci} provides consistent results, i.e.\ eigenvector components $v^j_m(t,t_0)$, which are in agreement with the coefficients $\tilde{v}^j_m$.


\subsection{\label{EQN856}Taking into account quark propagation within a timeslice}

When including quark propagation within a timeslice, correlation functions of the quark-antiquark interpolating field (Eq.\ (\ref{EQN002})) and the four-quark interpolating fields (Eqs.\ (\ref{EQN007}) to (\ref{EQN003})) are non-zero. Thus, all interpolating fields probe the same sector and the quark-antiquark interpolating field can be included in the analysis.

In Fig.\ \ref{fig:effective} we show effective energies obtained by solving the GEVP for two different correlation matrices:
\begin{figure}[htb]
\begin{center}
\includegraphics[width=8.2cm]{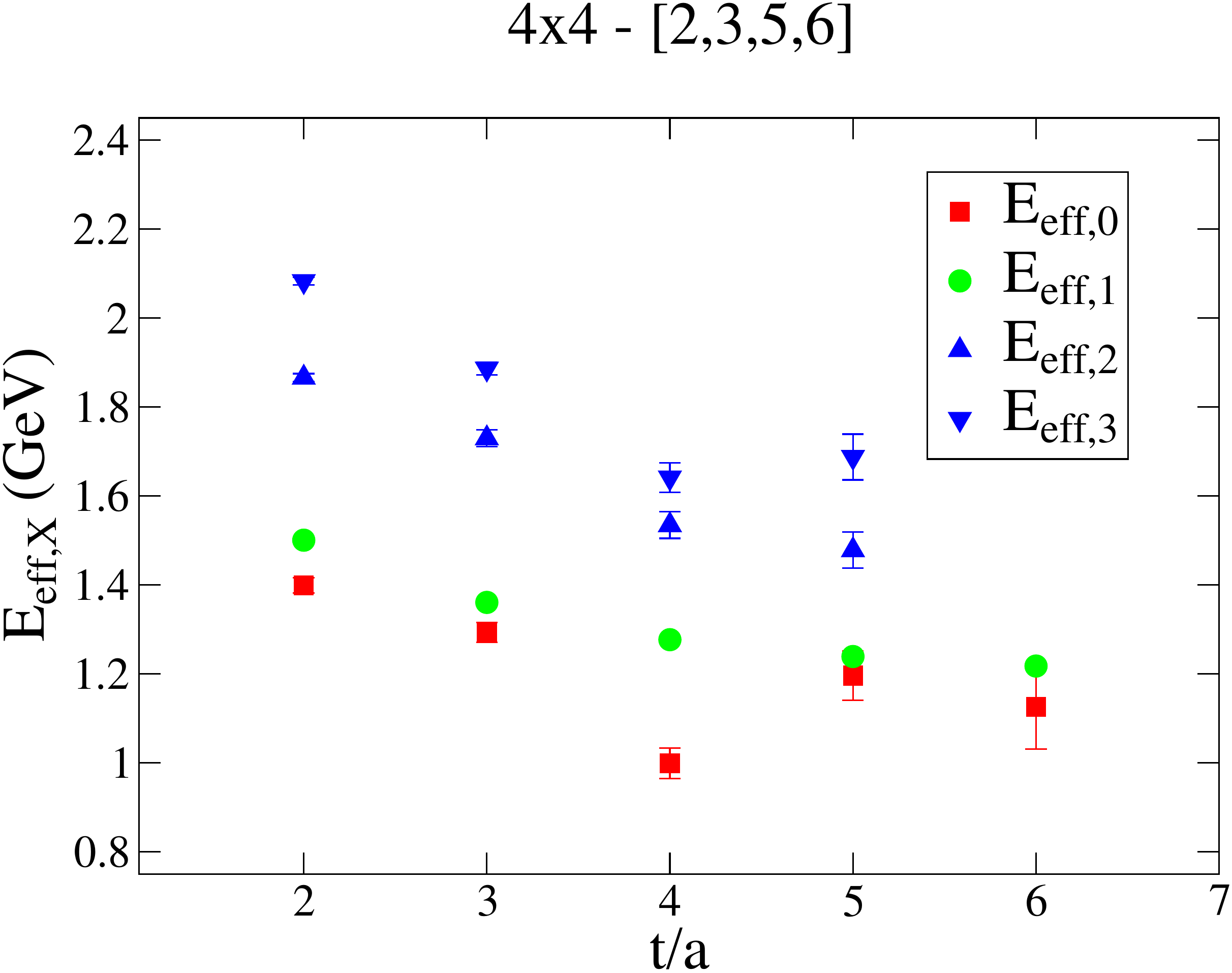} \\
\vspace{0.4cm}
\includegraphics[width=8.2cm]{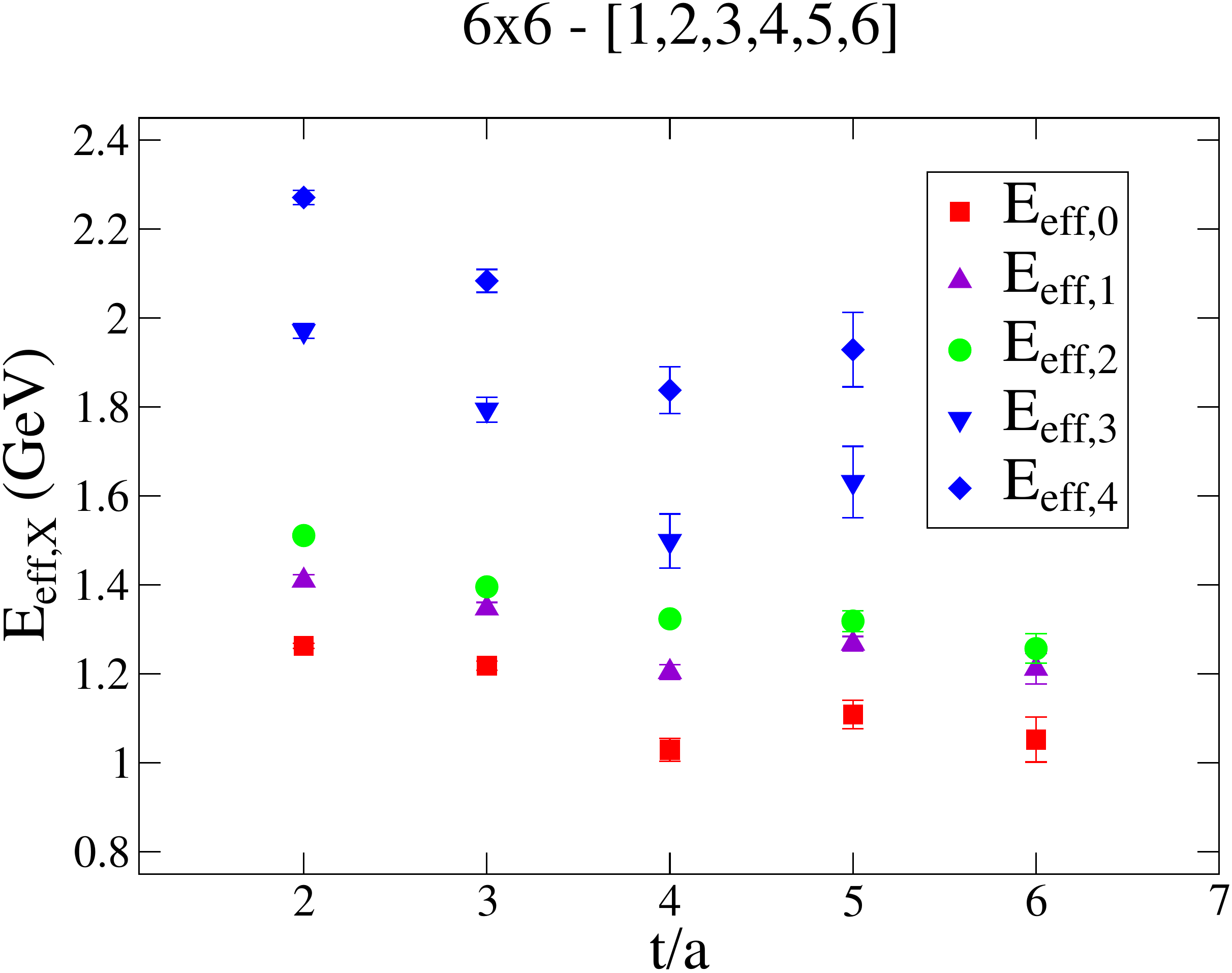}
\end{center}
\caption{\label{fig:effective}Effective energies obtained with the GEVP method (quark propagation within a timeslice taken into account). \textbf{(top)}~$4 \times 4$ correlation matrix with interpolating fields $\mathcal{O}^1 = \mathcal{O}^{q \bar q}$ and $\mathcal{O}^4 = \mathcal{O}^{Q \bar{Q}}$ excluded. \textbf{(bottom)}~Full $6 \times 6$ correlation matrix.}
\end{figure}
\begin{itemize}
\item $4 \times 4$ correlation matrix with interpolating fields $\mathcal{O}^2 = \mathcal{O}^{K \bar{K}, \ \textrm{point}}$, $\mathcal{O}^3 = \mathcal{O}^{\eta_{s} \pi, \ \textrm{point}}$, $\mathcal{O}^5 = \mathcal{O}^{K \bar{K}, \ \textrm{2part}}$ and $\mathcal{O}^6 = \mathcal{O}^{\eta_{s} \pi, \ \textrm{2part}}$, i.e.\ the quark-antiquark and the diquark-antidiquark interpolating fields excluded;

\item full $6 \times 6$ correlation matrix, i.e.\ interpolating fields $\mathcal{O}^1 = \mathcal{O}^{q \bar q}$ and $\mathcal{O}^4 = \mathcal{O}^{Q \bar{Q}}$ also considered.
\end{itemize}
In comparison to the effective energies from Fig.\ \ref{FIG005}, where quark propagation within a timeslice has been neglected, statistical uncertainties drastically increase. For a detailed discussion concerning the reason for this increase cf.\ e.g.\ \cite{Abdel-Rehim:2017dok}, section~4.4.3. The signal is essentially lost for temporal separations $t \gtapprox 7 \, a$. For temporal separations $t \ltapprox 6 \, a$ effects related to multiparticle states, as discussed in section~\ref{sec:euccor}, are negligible and, thus, can be ignored throughout this subsection. Due to the large statistical uncertainties, the identification of plateaus and energy differences is rather difficult. Nevertheless, there is a clear qualitative difference between the results from the $4 \times 4$ and the $6 \times 6$ correlation matrix. In the $4 \times 4$ plot there seem to be only two low-lying states around $1100 \, \textrm{MeV}$ to $1200 \, \textrm{MeV}$, while the next two states are significantly above, somewhere in the energy region of momentum excitations. Thus, the observed spectrum is consistent with the expected spectrum of two-meson states. In contrast to that, in the $6 \times 6$ plot, i.e.\ when taking also the quark-antiquark and the diquark-antidiquark interpolating fields into account, there is an additional third state around $1200 \, \textrm{MeV}$, which could correspond to the $a_0(980)$ meson. Similar plots have been obtained for $5 \times 5$ correlation matrices, where either the quark-antiquark interpolating field is considered, but not the diquark-antidiquark interpolating field, or vice versa. However, the effective energy of the additional state is somewhat larger and has larger statistical errors, when $\mathcal{O}^1 = \mathcal{O}^{q \bar q}$ is excluded. In summary, these results indicate that in addition to the expected two-meson states there is another low-lying state in the region of $1100 \, \textrm{MeV}$ to $1200 \, \textrm{MeV}$, which is predominantly of quark-antiquark type with a smaller, but non-negligible diquark-antidiquark component.

\begin{figure}[htb]
\begin{center}
\includegraphics[width=4.1cm,page=1]{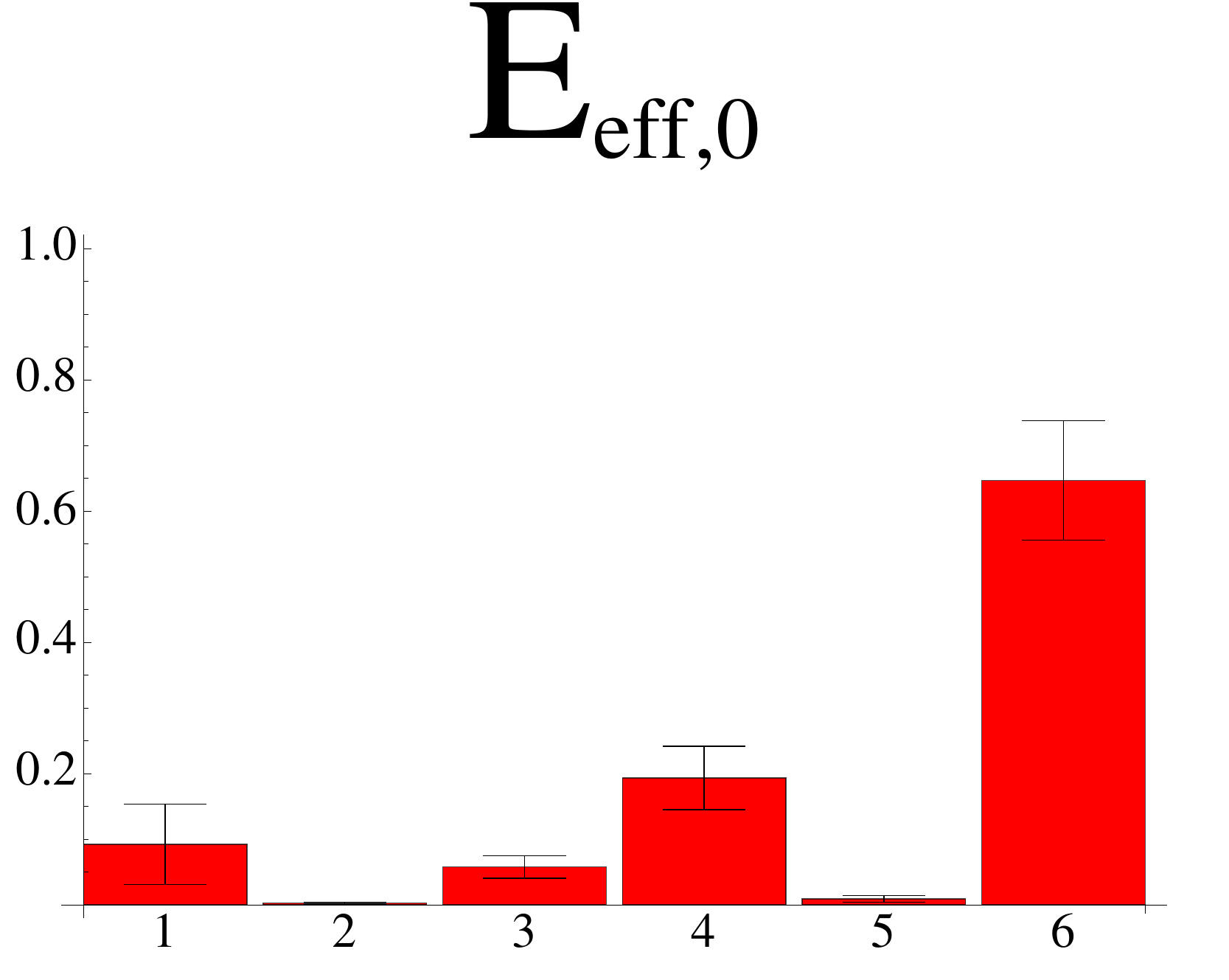}
\includegraphics[width=4.1cm,page=2]{figures/fig08.pdf} \\
\includegraphics[width=4.1cm,page=3]{figures/fig08.pdf}
\end{center}
\caption{\label{fig:effective2}Squared eigenvector components $(v^j_m(t=5a,t_0))^2$ for the three lowest states obtained with the GEVP method (quark propagation within a timeslice taken into account; full $6 \times 6$ correlation matrix).}
\end{figure}

This interpretation is confirmed by the squared eigenvector components $(v^j_m(t,t_0))^2$, which are plotted for the three lowest states in Fig.\ \ref{fig:effective2} for $t = 5 a$ (within statistical errors these eigenvector components are independent of $t$). The lowest state is mostly an $\eta + \pi$ two-meson state, the second-lowest state mostly a quark-antiquark state and the third-lowest state a $K + \bar{K}$ two-meson state. The lowest two states also contain a small diquark-antidiquark contribution.

\begin{figure}[htb]
\begin{center}
\includegraphics[width=8.2cm]{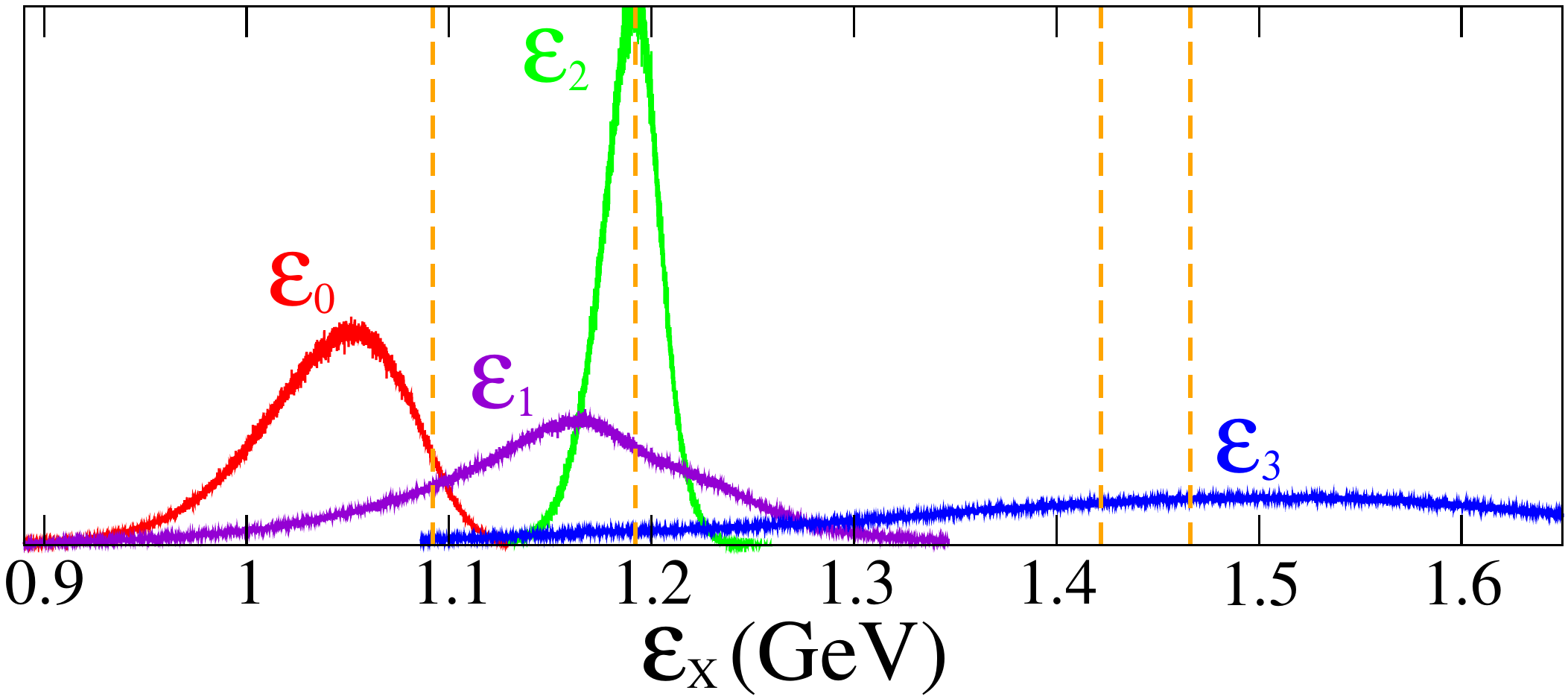} \\
\vspace{0.2cm}
\includegraphics[width=4.1cm,page=1]{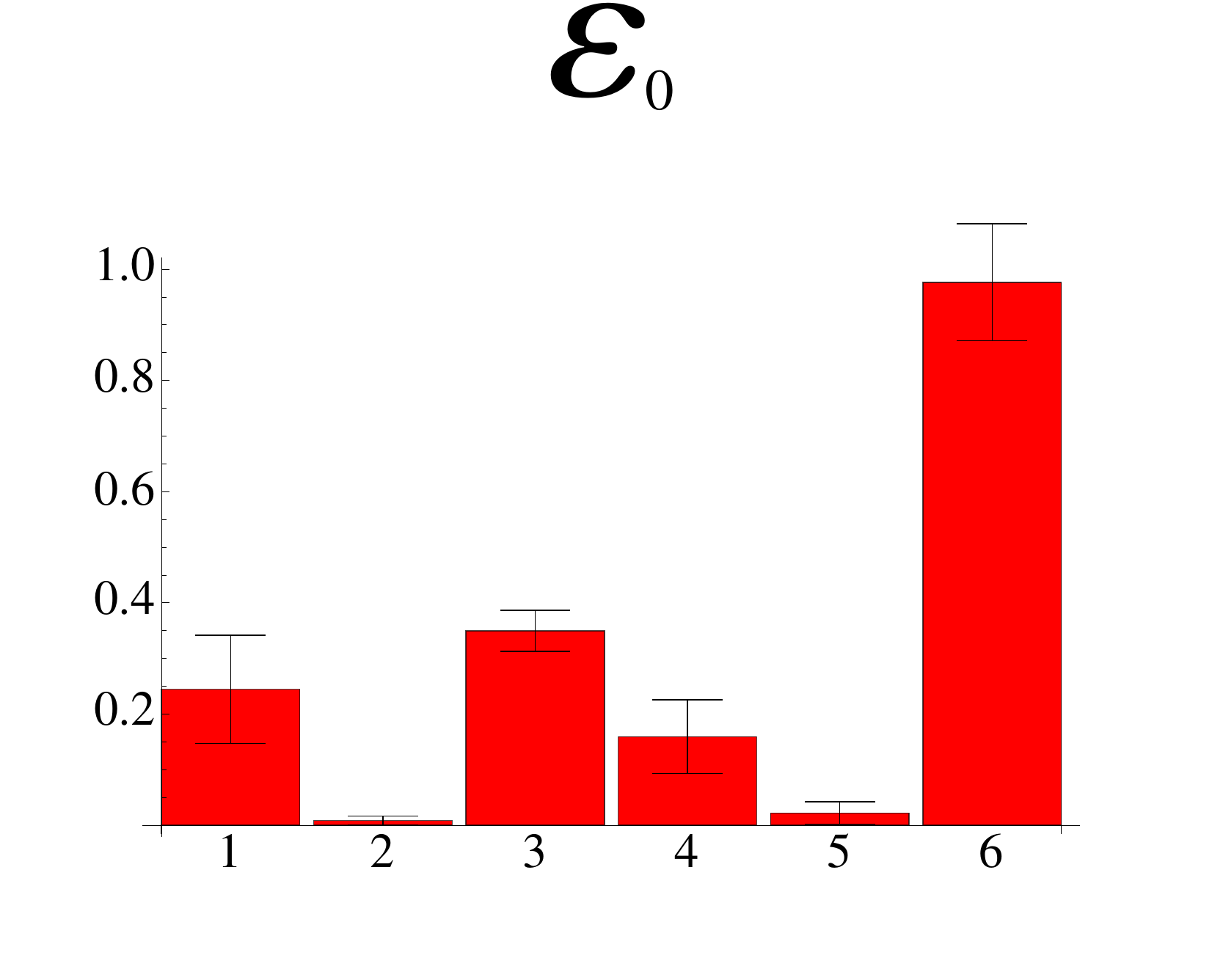}
\includegraphics[width=4.1cm,page=2]{figures/fig09b.pdf} \\
\includegraphics[width=4.1cm,page=3]{figures/fig09b.pdf}
\end{center}
\caption{\label{fig:AMIASwl}PDFs for energy differences and corresponding squared coefficients $(c^j_{\Omega,m})^2$ for the three lowest states obtained with the AMIAS method (quark propagation within a timeslice taken into account; full $6 \times 6$ correlation matrix).}
\end{figure}

To support these findings we have also determined energy differences from the full $6 \times 6$ correlation matrix with the AMIAS method. The corresponding PDFs for $t_\textrm{min} = 2 a$, $t_\textrm{max} = 8 a$ and eight terms in the truncated sum of the fit function (\ref{EQN006}) is shown in Fig.\ \ref{fig:AMIASwl}. There are two clear peaks consistent with the expected energies of two-particle $\eta + \pi$ and $K + \bar{K}$ states with both mesons at rest:
\begin{itemize}
\item $\mathcal{E}_0 = 1039(39) \, \textrm{MeV}$ \\ (expectation: $m_\pi + m_\eta \approx 1092 \, \textrm{MeV}$),

\item $\mathcal{E}_2 = 1192(11) \, \textrm{MeV}$ \\ (expectation: $2 m_K \approx 1192 \, \textrm{MeV}$).
\end{itemize}
Moreover, there is another peak signaling the existence of an additional third state in the same energy region with
\begin{itemize}
\item $\mathcal{E}_1 = 1124(76) \, \textrm{MeV}$,
\end{itemize}
i.e.\ significantly below the expectation for the lowest two-particle states with one quantum of relative momentum, $(m_\eta^2 + p_\textrm{min}^2)^{1/2} + (m_\pi^2 + p_\textrm{min}^2)^{1/2} \approx 1422 \, \textrm{MeV}$.

To clarify the structure of the extracted energy eigenstates, we also show the corresponding squared coefficients $(c^j_{\Omega,m})^2$, $m = 0,1,2$ in Fig.\ \ref{fig:AMIASwl} \footnote{Even though the coefficients $\tilde{v}^j_m$, $m = 0,1,2$ could be interpreted in a more straightforward way, their statistical errors are rather large, since their computation requires $c^j_{\Omega,m}$ not only for $m = 0,1,2$ but also for $m = 3,4,5$ (cf.\ eq.\ (\ref{EQN245})), where the latter have large statistical uncertainties.}. The two lowest states corresponding to the energy differences $\mathcal{E}_0$ and $\mathcal{E}_1$ contribute to the two-meson trial state $\mathcal{O}^6 | \Omega \rangle = \mathcal{O}^{\eta_{s} \pi, \ \textrm{2part}} | \Omega \rangle$ as well as to the quark-antiquark and the diquark-antidiquark trial states $\mathcal{O}^1 | \Omega \rangle = \mathcal{O}^{q \bar q} | \Omega \rangle$ and $\mathcal{O}^4 | \Omega \rangle = \mathcal{O}^{Q \bar Q} | \Omega \rangle$. The second excitation corresponding to the energy difference $\mathcal{E}_2$ contributes almost exclusively to the two-meson trial state $\mathcal{O}^5 | \Omega \rangle = \mathcal{O}^{K \bar{K}, \ \textrm{2part}} | \Omega \rangle$. These AMIAS results are in agreement with the GEVP results discussed above and, thus, confirm our previous interpretation that there is an additional state in the energy region of $1100 \, \textrm{MeV}$ to $1200 \, \textrm{MeV}$, which could correspond to the $a_0(980)$ meson. This additional state has both a quark-antiquark and a diquark-antidiquark component.

\begin{figure}[htb]
\begin{center}
\includegraphics[width=8.2cm]{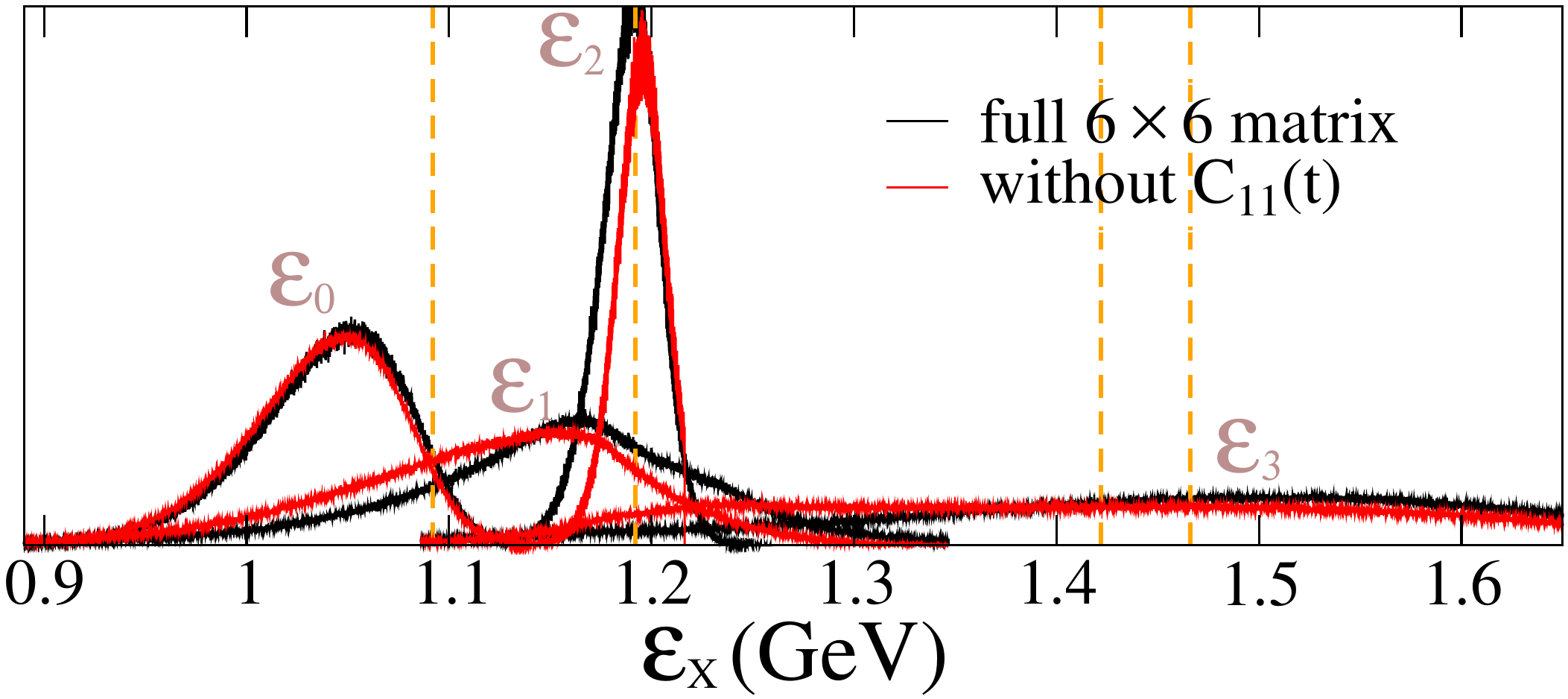}
\end{center}
\caption{\label{fig:AMIAS3}PDFs for energy differences obtained with the AMIAS method (quark propagation within a timeslice taken into account; full $6 \times 6$ correlation matrix and the same matrix without $C_{1 1}(t)$).}
\end{figure}

One of the advantages of the AMIAS method, compared to e.g.\ the GEVP method, is that one can use an arbitrary selection of correlation matrix elements for an analysis. To check the correctness and stability of our results, in particular the existence of an additional low-lying state with significant quark-antiquark component, we compare the PDFs for energy differences based on two different analyses and sets of correlation matrix elements in Fig.\ \ref{fig:AMIAS3}:
\begin{itemize}
\item[(a)] the full $6 \times 6$ correlation matrix (same data as in Fig.\ \ref{fig:AMIASwl}),

\item[(b)] as in (a), but the diagonal element $C_{1 1}(t)$ is excluded from the analysis, i.e.\ $j = k = 1$ is omitted in the sum $\sum_{j,k}$ in Eq.\ (\ref{eq:chi2}); this implies that the correlation of the quark-antiquark interpolating field $\mathcal{O}^1 = \mathcal{O}^{q \bar q}$ with itself is excluded, while correlations with the other four-quark interpolating fields are still included.
\end{itemize}
Fig.\ \ref{fig:AMIAS3} represents an important check of our results and confirms our interpretation. The additionally observed state is not just generated by adding an essentially independent interpolating field $\mathcal{O}^1 = \mathcal{O}^{q \bar q}$. This quark-antiquark interpolating field $\mathcal{O}^1$ couples to the four-quark interpolating fields $\mathcal{O}^2$ to $\mathcal{O}^6$ and the additional low-lying state can be clearly observed, even when $C_{1 1}(1)$ is excluded from the analysis.

\section{\label{sec:conclusions}Summary and conclusions}

We have computed the low-lying spectrum of the $I(J^P) = 1(0^+)$ sector using lattice QCD. To this end, we have considered a variety of interpolating fields representing quark-antiquark and four-quark bound states as well as two-meson scattering states. In addition to the expected two-meson scattering states we found another state in the energy region of $1100 \, \textrm{MeV}$ to $1200 \, \textrm{MeV}$, which is a candidate for the $a_0(980)$ meson. This state is predominantly generated by the quark-antiquark interpolating field, but also receives sizable contributions from the diquark-antidiquark interpolating field, i.e.\ a likely interpretation is that it is mainly a quark-antiquark state with a minor tetraquark component. To some extent this is supported by a computation, where we have neglected quark propagation within a timeslice. Then the quark-antiquark interpolating field decouples from the four-quark interpolating fields and an analysis of the four-quark correlation matrix only yields the expected two-meson scattering states.

We have cross-checked and confirmed our results by applying two different analysis methods, the well-known GEVP method and the rather new AMIAS method. The latter is particularly suited to handle lattice QCD data with large uncertainties, because it also exploits correlation functions at small temporal separations.

A current source of systematic error, which we plan to eliminate in the future, is the unphysically large $u$ and $d$ quark masses corresponding to $m_\pi \approx 300 \, \textrm{MeV}$. A resonance study of $a_0(980)$ using L\"uscher's finite volume approach was presented in Ref.\ \cite{Dudek:2016cru} using, however, $u$ and $d$ quark masses corresponding to $m_\pi \approx 391 \, \textrm{MeV}$. In Ref.\ \cite{Guo:2016zep}, which is an analysis of the same lattice QCD data using chiral effective field theory, it was pointed out that lighter $u$ and $d$ quark masses are essential to obtain more precise results. However, the extension of the method to lighter $u$ and $d$ quark masses is very demanding, since it requires the identification of all energy states below that of the $a_0(980)$. Therefore, the work presented here constitutes an important preparatory step.

Another obvious direction to extend this research is to investigate the $D_{s0}^\ast(2317)$ meson, which is also considered as a tetraquark candidate. Our techniques and code can be used with only minor changes, since both $a_0(980)$ and $D_{s0}^\ast(2317)$ have the same quantum numbers $J^P = 0^+$ and the same flavor structure, i.e.\ a quark-antiquark pair of different flavor $\bar{q}_1 q_2$ and possibly another quark-antiquark pair of the same flavor $\bar{q}_3 q_3$. Such an investigation could be of particular interest, because lattice QCD studies like those presented in \cite{Moir:2013ub,Kalinowski:2015bwa,Cichy:2016bci}, which do not include four-quark interpolating fields, found masses significantly above the experimental result. Other lattice QCD studies, e.g.\ \cite{Lang:2014yfa,Bali:2017pdv}, which include four-quark interpolating fields found a state below the $D K$ threshold much closer to the experimental result. Thus, it would be interesting to further explore the existence and mass of a $D_{s_0}^*(2317)$ state below the $D K$ threshold, by varying the set of interpolating fields considered in the analysis, and to investigate its internal structure, e.g.\ by employing a variety of tetraquark interpolating fields.


\section{Acknowledgments}

We would like to thank A. Abdel Rehim for his contributiona at an early stage of this work.

This work was cofunded by the European Regional Development Fund and the Republic of Cyprus through the Research Promotion Foundation (Project Cy-Tera $\textrm{Υ}\Pi\textrm{O}\Delta\textrm{OMH}/\Sigma\textrm{TPATH}/0308/31$) by the grant cypro914. This work was supported in part by the Helmholtz International Center for FAIR within the framework of the LOEWE
program launched by the State of Hesse.

J.B.~and M.W.~acknowledge support by the Emmy Noether Programme of the DFG (German Research Foundation), grant WA
3000/1-1. M.D.B.~is grateful to the Theoretical Physics Department at CERN for the hospitality and support. 

All calculations are performed on the LOEWE-CSC high-performance computer of the Frankfurt University and on Cy-Tera at the Computation-based Science and Technology Research Center of The Cyprus Institute. We would like to thank HPC-Hessen, funded by the State Ministry of Higher Education, Research and the Arts, for programming advice. Computations have been performed using the Chroma software library \cite{Edwards:2004sx} with a
multigrid solver \cite{Babich:2010qb}.


\bibliography{a0}


\end{document}